\documentclass[english,aps,prl,amsmath,amssymb,twocolumn,letterpaper,showpacs,superscriptaddress]{revtex4-1}

\usepackage{mathtools} %floor and ceiling functions

\usepackage[normalem]{ulem} %Strikethrough

\usepackage{amsmath}
\usepackage{physics}
\usepackage{graphicx}
\usepackage{babel}
\usepackage{amstext}
\usepackage{esint}
\usepackage[unicode=true,pdfusetitle,
bookmarks=true,bookmarksnumbered=false,bookmarksopen=false, breaklinks=false,pdfborder={0 0 1},backref=false,colorlinks=false]
 {hyperref}
\usepackage{pdfpages}	% Packages needed
\usepackage{pgffor}		% to include supplement
\usepackage{upgreek}
\usepackage{braket}

\newcommand{\bk}{\mathbf{k}}
\newcommand{\br}{\mathbf{r}}
\newcommand{\mum}{\upmu\mathrm{m}}

\makeatletter
\AtBeginDocument{\let\LS@rot\@undefined} %This fixes the issue
\makeatother							 %that pdfpages rotates the pages by 90deg.

\begin{document}

\title{Influence of disorder on antidot vortex Majorana states in 3D topological insulators }

\author{Rafa\l{} Rechci\'nski}
\affiliation{Institute of Physics, Polish Academy of Sciences, Aleja Lotnikow 32/46, PL-02668 Warsaw, Poland}
\affiliation{International Research Centre MagTop, Institute of Physics, Polish Academy of Sciences, Aleja Lotnikow 32/46, PL-02668 Warsaw, Poland}
\affiliation{Microsoft Quantum, Station Q, University of California, Santa Barbara, California 93106, USA}

\author{Aleksei Khindanov}
\affiliation{Department of Physics, University of California, Santa Barbara, California 93106, USA}
\affiliation{Ames National Laboratory, U.S. Department of Energy, Ames, Iowa 50011, USA}
\author{Dmitry I. Pikulin}
\affiliation{Microsoft Quantum, Station Q, University of California, Santa Barbara, California 93106, USA}
\affiliation{Microsoft Quantum, Redmond, Washington 98052, USA}

\author{Jian Liao}
\affiliation{Department of Physics and Astronomy, Purdue University, West Lafayette, Indiana 47907, USA}

\author{Leonid P. Rokhinson}
\affiliation{Department of Physics and Astronomy, Purdue University, West Lafayette, Indiana 47907, USA}
\affiliation{Elmore Family School of Electrical and
Computer Engineering, Purdue University, West Lafayette,
Indiana 47907, USA}
\affiliation{Birck Nanotechnology Center, and Purdue
Quantum Science and Engineering Institute, Purdue
University, West Lafayette, Indiana 47907, USA}

\author{Yong P. Chen}
\affiliation{Department of Physics and Astronomy, Purdue University, West Lafayette, Indiana 47907, USA}
\affiliation{Elmore Family School of Electrical and
Computer Engineering, Purdue University, West Lafayette,
Indiana 47907, USA}
\affiliation{Birck Nanotechnology Center, and Purdue
Quantum Science and Engineering Institute, Purdue
University, West Lafayette, Indiana 47907, USA}

\author{Roman M. Lutchyn}
\affiliation{Microsoft Quantum, Station Q, University of California, Santa Barbara, California 93106, USA}

\author{Jukka I. V\"{a}yrynen}
\affiliation{Department of Physics and Astronomy, Purdue University, West Lafayette, Indiana 47907, USA}

\date{\today}
\begin{abstract}
Topological insulator/superconductor two-dimensional heterostructures are promising candidates for realizing topological superconductivity and Majorana modes. In these systems, a vortex pinned by a prefabricated antidot in the superconductor can host Majorana zero-energy modes (MZMs), which are exotic quasiparticles that may enable quantum information processing. However, a major challenge is to design devices that can manipulate the information encoded in these MZMs. One of the key factors is to create small and clean antidots, so the MZMs, localized in the vortex core, have a large gap to other excitations. If the antidot is too large or too disordered, the level spacing for the subgap vortex states may become smaller than temperature. In this paper, we numerically investigate the effects of disorder, chemical potential, and antidot size on the subgap vortex spectrum, using a two-dimensional effective model of the topological insulator surface. Our model allows us to simulate large system sizes with vortices up to $1.8\, \mum$ in diameter (with a $6\,\mathrm{nm}$ lattice constant). We also compare our disorder model with the transport data from existing experiments. We find that the spectral gap can exhibit a nonmonotonic behavior as a function of disorder strength, and that it can be tuned by applying a gate voltage.
\end{abstract}

\maketitle

\section{Introduction}
Majorana zero modes (MZMs) are exotic quasiparticles that obey non-Abelian exchange statistics and can be used for topological quantum computation~\cite{Kitaev01,RevModPhys.80.1083,Alicea12a,10.21468/SciPostPhys.6.5.055,dassarma2015}. Topological protection of qubits based on MZMs is governed by the bulk excitation (topological) gap $\Delta_0$. The probability of errors induced by local noise sources in such qubits is suppressed as $\exp(-\Delta_0/T)$ and $\exp(-\Delta_0 L/v)$ with $T$, $L$, and $v$ being temperature, distance between MZMs, and Fermi velocity, respectively. 
Another important energy scale for Majorana-based qubits is the minigap $\Delta_m$, which is the energy difference between the zero-energy states and the higher-energy localized states. For qubit operation, the minigap needs to be larger than the temperature to enable fast and reliable measurement of the MZM parity~\cite{Akhmerov2010}.

Presently, the most commonly investigated platform for creating MZMs is semiconductor nanowires with large spin-orbit coupling and proximity-induced superconductivity~\cite{Lutchyn10,Oreg10}.
Despite strong experimental signatures of the topological phase in these nanowires reported to date~\cite{2022arXiv220702472A,PhysRevB.107.245304}, a large magnetic field necessary to drive the nanowire across the topological phase transition poses significant engineering challenges. These include disorder enabled by time-reversal symmetry breaking, the suppression of the superconductivity in the parent superconductor, and the emergence of the orbital effect due to the magnetic field~ \cite{Alicea12a,Lutchyn17,Nijholt2016,Winkler2019,2021NatRM...6..944F}.

On the other hand, an MZM platform based on the surface of a three-dimensional topological insulator (3D TI) covered by a superconductor (SC)~\cite{PhysRevLett.100.096407} does not require a large magnetic field, as the surface of the 3D TI naturally realizes a topological superconducting state that can host MZMs in vortices~\cite{2010RvMP...82.3045H,Breunig2021,Proximity2022}.
This eliminates the challenges posed by the large magnetic field and makes the 3D TI-SC platform potentially more attractive for realizing topological superconductivity.
To date, a number of studies have reported the observation of induced superconductivity in 3D TI-based Josephson junctions~\cite{2011NatCo...2..575S,2012NatMa..11..417V,2012PhRvL.109e6803W,2015PhRvL.114f6801S,2013arXiv1307.7764K,2017NatCo...8.2019C,Jauregui2018,2018NanoL..18.5124G,2018NatCo...9.3478C,PhysRevMaterials.4.094801,2021arXiv211001039R}.

A possible way to create and control vortices in a 3D TI-SC platform is to use an antidot structure, where part of the SC is removed, as shown in Fig.~\ref{fig:sample_sketch}(a). The size of the antidot can be chosen such that a small magnetic field can induce a vortex with a single MZM. For a large antidot, the magnetic flux quantum required for a MZM can then be achieved with a relatively small magnetic field. This should be contrasted with recently studied zero modes in  Abrikosov vortices at high magnetic fields in Fe-based type-II superconductors (typically a few  Tesla)~\cite{2019NatPh..15.1181K,doi:10.1126/sciadv.aay0443} or in proximitized topological insulator surfaces (at fields of order $0.1\,\mathrm{T}$)~\cite{PhysRevLett.114.017001,PhysRevLett.116.257003}. Moreover, the electron density inside the antidot can be tuned by a gate voltage, since it is not screened by the SC~\cite{lee2020scalable}. Despite the advantages of this platform, there are still many open questions and challenges that need to be addressed. Most of the previous studies on the 3D TI antidot structure have focused on either the clean~\cite{PhysRevB.89.085409,Deng2020,Ziesen21} or the strongly disordered~\cite{PhysRevB.86.035441} regimes, where analytical results can be obtained. However, these regimes may not be relevant for realistic experiments, where intermediate disorder strengths and finite system sizes are more common. Therefore, numerical simulations are needed to provide more accurate predictions and guidance for experimental works regarding disorder requirements. 

Recently, Ref.~\cite{PhysRevLett.130.106001} performed a numerical study of the low-energy antidot subgap spectrum using an effective continuous model that treats the SC outside the antidot as a boundary condition for the 3D TI surface and incorporates a Gaussian-distributed disorder potential through random matrix elements. Furthermore, the article explored the possibility of generating an ensemble of disorder profiles in a single sample, by means of several electrostatic finger gates distributed throughout the antidot area.  

In this paper, we present a more microscopic modeling of the antidot structure using a two-dimensional (2D) effective lattice model for the proximitized 3D TI surface, with the disorder potential included explicitly. The model is described in detail in Sec.~\ref{sec:model}. It allows us to simulate large system sizes up to $3.6\, \mum \times 3.6\, \mum$ with a lattice spacing $a = 6\, \mathrm{nm}$, and to capture the low-energy physics near the Dirac point. We investigate how the minigap depends on various parameters, such as disorder strength, antidot radius, and chemical potential (i.e., electron density), in Sec.~\ref{sec:results}. We also compare our disorder model with experimental mobility data to estimate realistic disorder levels in existing materials. We discuss the implications of our results for the feasibility of observing MZMs in this platform in Sec.~\ref{sec:conclusions}.

\section{Model \label{sec:model} }
\subsection{Effective model of the proximitized TI surface}

To be able to efficiently simulate a 2D surface of a 3D TI with a single Dirac cone, we utilize an effective square-lattice model of the surface that breaks global time-reversal symmetry~\cite{Pikulin2017,PhysRevB.86.155146}. Within this model, the Hamiltonian of a uniform surface has the form
\begin{equation} \label{eq:TI_surf_Hamiltonian}
    h_\mathrm{TI}(\bk) = \lambda(s_x \sin a k_x + s_y \sin a k_y) + M_\bk s_z - \mu,
\end{equation}
where $M_\bk = m \bigl(\frac{3}{2} - \cos a k_x - \cos a k_y + \frac{1}{4} \cos 2 a k_x + \frac{1}{4} \cos 2 a  k_y \bigr)$ is the time-reversal symmetry breaking term, $s_i$ are Pauli matrices in the spin space, $k_i$ are electronic momenta, $a$ is the lattice spacing, and $\mu$ is the chemical potential. 
In the absence of $M_\bk$, the Hamiltonian has four Dirac cones at the high symmetry points 
$\Gamma = \left(0,0\right)$, $\mathrm{X} = \left(\frac{\pi}{a},0\right)$, $\mathrm{Y} = \left(0,\frac{\pi}{a}\right)$, and $\mathrm{M} = \left(\frac{\pi}{a},\frac{\pi}{a}\right)$; the model parameter $\lambda$ determines the Dirac velocity, $v_D = \lambda a/\hbar$. The term $M_\bk$ breaks the global time-reversal symmetry and opens gaps of the order of $|m|$ at all high symmetry points except for $\Gamma$, where its effect on the Dirac spectrum is minor, given that $M_\bk \approx \frac{m a^4}{8}(k_x^4+k_y^4)$ for small $\bk$. 
The presence of the term $M_\bk$ thus effectively creates a 2D surface with a single Dirac cone at $\Gamma$.
We refer the reader to Ref.~\cite{Pikulin2017} for a detailed description of the model.

Proximity-induced superconductivity is included by constructing a Bogoliubov--de Gennes Hamiltonian $H=\frac{1}{2}\sum_\bk \Psi^\dagger_\bk H_\mathrm{BdG}(\bk)\Psi_\bk$ with
\begin{equation} \label{eq:BgD_Hamiltonian}
    H_\mathrm{BdG}(\bk) = \begin{pmatrix}
     h_\mathrm{TI}(\bk) & i \Delta s_y \\
     -i \Delta^* s_y & -h^*_\mathrm{TI}(-\bk) \\
    \end{pmatrix},
\end{equation}
where $\Delta$ is the proximity-induced effective pairing potential in the TI and the basis spinor is $\Psi_\bk^\dagger=(c^\dagger_{\uparrow,\bk},c^\dagger_{\downarrow,\bk},c_{\uparrow,-\bk}, c_{\downarrow,-\bk}  )$. 
The superconducting correlations in the TI are generated through the tunneling between the TI and the parent SC, and the effective Hamiltonian \eqref{eq:BgD_Hamiltonian} can be calculated by integrating out the SC degrees of freedom in the combined SC-TI system and computing the SC-TI interface self-energy~\cite{2010PhRvB..82i4522S,Sedlmayr2021}.
$\Delta$ obtained in this way depends on the pairing potential in the parent SC as well as on the strength of the TI-SC coupling, and is in general a function of the spatial coordinate $\br$. For ease of analysis, we treat $\Delta$ as a free parameter---this common simplification does not alter the conclusions of our paper.

In our simulations we choose a discretization with the lattice spacing $a=0.03 \pi \xi_0$, where $\xi_0 = \hbar v_D/(\pi \Delta_0) $ is the superconducting coherence length \footnote{\label{ft:pixi}The decay length of the wave functions in the SC is given by $ \pi \xi_0 $. Based on assumption $\hbar v_D = 200\,\mathrm{meV}\,\mathrm{nm}$ and $\Delta_0=1\,\mathrm{meV}$, for conversion to physical units we use $\pi \xi_0 = 200\,\mathrm{nm}$. However, our analysis is independent of the precise value of $\xi_0$.} in a clean proximitized system, and $\Delta_0$ denotes the induced SC gap. By definition, $\lambda = \pi \Delta_0 \xi_0/a$. Furthermore, we fix $m=-1.5 \lambda$ (the choice of sign is arbitrary). As demonstrated in Appendix~\ref{app:validity}, up to energies $|E|\ll 60 \Delta_0$ the term linear in $\bk$ in the power series expansion of~\eqref{eq:TI_surf_Hamiltonian} dominates over higher order terms. Therefore, in the energy range investigated in this paper ($|E| \leq 15\Delta_0$), our lattice model is a good approximation of the TI surface Dirac Hamiltonian. 

We note that, when $|\Delta_0|\ll |m|$ and  $|\mu|\ll |m|$, the model has a non-zero Chern number $C=\mathrm{sign}(m)$, which is due to the presence of time-reversal symmetry breaking terms in~\eqref{eq:BgD_Hamiltonian}. Hence, our finite-size system with open boundary conditions will feature a chiral edge mode, which is an artifact of the effective model. Effectively, the edge of the sample is analogous to a boundary between a SC-proximitized TI surface and a magnetic-insulator-proximitized TI surface with Chern number $C=-\mathrm{sign}(m)$. In the following simulations, the edges are located at least $12 \xi_0$ away from the antidot boundary. This ensures that the spurious edge states have negligible influence on the antidot spectrum. 

\subsection{Model of the antidot system}
To simulate the antidot device, we write the Hamiltonian~\eqref{eq:BgD_Hamiltonian} in the position space, and allow spatial variation of the chemical potential $\mu$ and the pairing potential $\Delta$. Spatial variation of $\lambda$, which could emerge due to the proximity-induced renormalization of $v_D$, is neglected.

\begin{figure}
\centering
\includegraphics[width=1.0\columnwidth]{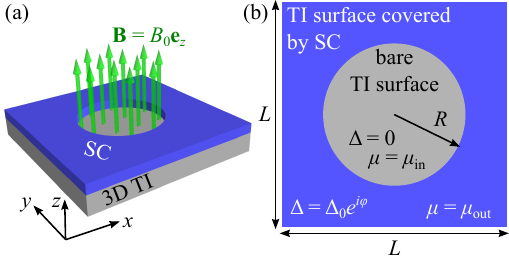}
\caption{(a) Sketch of the sample structure and the modeled distribution of the magnetic field. (b) Top view of the sample with a schematic representation of the parameters of the model.}
 \label{fig:sample_sketch}
\end{figure}

A schematic sketch of the modeled sample is shown in Fig.~\ref{fig:sample_sketch}. The system comprises a square fragment of a 3D TI surface with a side length of $L=3.6\,\mum \approx 56.5 \xi_0$, covered by a SC layer everywhere except for a circular antidot area of radius $R$ in the middle. We fix the coordinate system origin in the center of the antidot and denote $(r,\varphi)$ as polar coordinates. We assume that the magnetic field is present exclusively in the antidot area and has the form
\begin{equation}
    \mathbf{B}(\br) =
    \begin{cases}
        (0,0,B_0)   & \textrm{for } r<R, \\
        (0,0,0)   & \textrm{for } r\geq R, \\
    \end{cases}
\end{equation}
which is a valid approximation in the case of a thick superconductor with a short London penetration length. We choose $B_0 = \Phi_0/(\pi R^2)$, such that the sample is permeated by a single flux quantum $\Phi_0=h/(2|e|)$. The magnetic vector potential compatible with the assumed magnetic field distribution is expressed in the London gauge as
\begin{equation}
 \mathbf{A}(\br) = 
\begin{dcases}
    \frac{\Phi_0}{2\pi R^2 } (-y,x,0) & \textrm{for } r<R, \\
    \frac{\Phi_0}{2\pi r^2} (-y,x,0) & \textrm{for } r\geq R. \\
    \end{dcases}.
\end{equation}
The flux trapped by the antidot induces a phase winding of the pairing potential
\begin{equation}
    \Delta(\br) = \begin{cases}
    0 & \textrm{for } r<R, \\
    \Delta_0 e^{-i \varphi} & \textrm{for } r\geq R, \\
    \end{cases}.
\end{equation}
Furthermore, the vector potential $\mathbf{A}(\br)$ is introduced into the hopping terms via the Peierls substitution,
\begin{equation}
    c^\dagger_{\br_1} c_{\br_0} \rightarrow 
    c^\dagger_{\br_1} c_{\br_0} \exp\left( -i\frac{|e|}{\hbar} \int_{L_{01}} \mathbf{A}(\br) \cdot d \br \right),
\end{equation}
where $L_{01}$ is the straight line pointing from $\br_0$ to $\br_1$.

The chemical potential profile is made up of two parts: $\mu(\br)=\mu_0(\br) + \delta\mu (\br)$. The first part describes an idealized impurity-free sample, while the second part represents the effect of the disorder. We assume that the chemical potential is fixed in the SC part of the system, while in the antidot it can be controlled by gating, and write
\begin{equation}
    \mu_0(\br) = \begin{cases}
    \mu_\textrm{in,0} & \textrm{for } r<R, \\
    \mu_\textrm{out} & \textrm{for } r\geq R. \\
    \end{cases}
\end{equation}

\subsection{Disorder model and its relation to the scattering rate }
The disorder in the system is modeled by distributing $2 N_C$ charged impurities at randomly chosen distinct lattice sites $\br^C_n$ throughout the sample. The charges have equal magnitude, although half of them are negative and half are positive, such that the net charge in the system is exactly zero, and the correction to the chemical potential due to disorder is
\begin{equation} \label{eq:delta_mu}
    \delta \mu (\br) = \sum_{n=1}^{N_C } V(\br-\br^C_n) - \sum_{n=N_C+1}^{2 N_C} V(\br-\br^C_n).
\end{equation}
As our TI surface model~\eqref{eq:TI_surf_Hamiltonian} features spin-polarized bands away from $\Gamma$, the disorder-induced scattering could potentially generate a magnetic gap in the surface Dirac spectrum. To avoid this issue, we choose the single impurity potential $V(\br)$ to have a Gaussian profile, such that the high momentum scattering is suppressed. The model potential in~\eqref{eq:delta_mu} is assumed to have the form
\begin{equation} \label{eq:imp_potential}
    V(\br) = -\frac{V_0}{\sqrt{\mathcal{N}} } \exp \left( -\frac{\br^2}{2 \sigma^2} \right),
\end{equation}
where $V_0$ is the magnitude of the potential, while $\sigma$ gives the radius of the potential well, and $  \mathcal{N} =  \sum_{i\in \textrm{lattice}} \exp [-\br_i^2/\sigma^2] $
is the normalization factor designed to fix the variance of $V(\br)$ at the lattice sites $\br_i$ such that $\sum_{i \in \textrm{lattice} } V^2(\br_i)= V_0^2$. The parameter $\sigma^{-1}$ gives an estimate for the largest momentum change in the scattering process. We fix $\sigma = 2.5 a$ in all calculations, for which choice the sum can be approximated by an integral and $\mathcal{N} \approx \pi (\sigma/a)^2$.

Finally, we connect the abstract parameters of the numerical model with measurable disorder characteristics. For $\mu_0 = 0$, we estimate the electron elastic scattering rate, averaged over fluctuations of $\delta \mu$, to be
\begin{equation} \label{eq:scatter_rate}
    \Gamma = \left\langle \frac{\hbar}{ \tau_\bk} \right\rangle_{\delta \mu} =    \sqrt{8 \pi}  \frac{\sigma^2 s^3}{\hbar^2 v_D^2}    \left( 1 - 2 \frac{\sigma^2 s^2}{\hbar^2 v_D^2}   \right),
\end{equation}
where 
\begin{equation}
    s^2 = \rho_\mathrm{imp} \frac{V^2_0 \pi\sigma^2 }{\mathcal{N}} \approx \rho_\mathrm{imp} V^2_0 a^2 
\end{equation} is the estimated variance of $\delta \mu(\br)$ at fixed $\br$. Expression~\eqref{eq:scatter_rate} is valid if $|\delta \mu| \ll |\hbar v_D/\sigma|$ everywhere in the sample. For stronger disorder, and for $\mu_0\neq 0$, we evaluate $\Gamma$ numerically. See Appendix~\ref{app:scattering} for the derivations of both the approximate and the numerical approaches and a plot of the dependence $\Gamma(s)$. Finally, we estimate the electron mean free path as
\begin{equation} \label{eq:mfp}
    l = v_D \left\langle \frac{1}{ \tau_\bk} \right\rangle_{\delta \mu}^{-1}  = \frac{\hbar v_D }{ \Gamma}.
\end{equation}

\section{Results \label{sec:results} }

Our main objective is to investigate how the minigap and the local density of states change upon introducing disorder in the sample. To that end, we calculated the energy spectrum of the antidot system, while varying the impurity potential magnitude $V_0$ and the impurity density $\rho_\mathrm{imp}$. In addition, we considered different values of the antidot radius $R$ and of the chemical potential $\mu_\mathrm{in,0}$ inside the sample, as these are the degrees of freedom that can be controlled directly in experiments. While, in principle, the results depend on the details of the distribution of impurities, in the main text we focus on the data calculated for a specific distribution, for which the minigap approaches zero at relatively weak disorder strengths, especially if the antidot radius equals about $14\xi_0$. With this method, we capture the qualitative features of the investigated phenomena that are likely to manifest in experimental observations of actual samples. Furthermore, to gain a more comprehensive view of the quantitative details, we calculated statistical distributions of the minigap value for various configurations of the antidot systems, with different realizations of random disorder. We discuss these results in Appendix~\ref{app:stats}. A thorough study of statistical distributions of energy levels in disordered antidot systems in the limit $\xi_0 \rightarrow 0 $ can be found in Ref.~\cite{PhysRevLett.130.106001}.

We characterize the antidot system with dimensionless quantities $\mu/\Delta_0$, $R/\xi_0= \pi R \Delta_0/(\hbar v_D)$, and $V_0/\Delta_0$. Furthermore, we fix the chemical potential in the SC part of the system to be $\mu_\mathrm{out}= 2 \Delta_0$. We have verified numerically that varying $\mu_\mathrm{out}$ in the range of several $\Delta_0$ has a negligible effect on the minigap, although the energies of higher excited states can be affected more significantly.

\begin{figure*}
\centering
\includegraphics[scale=1]{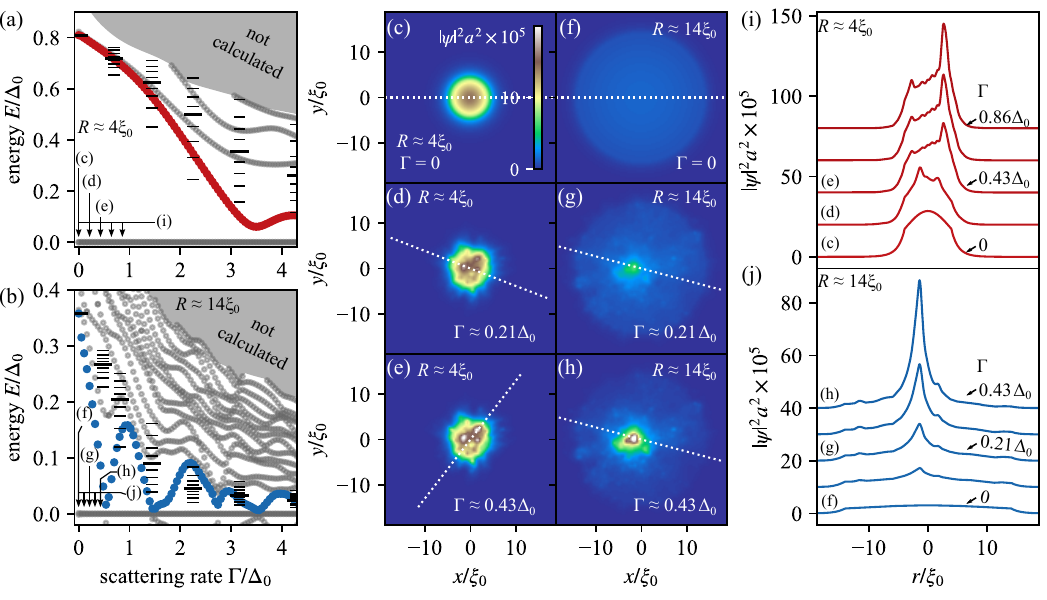}
\caption{
\textbf{Left column:} Energy spectra of the vortex cores pinned inside the antidots of radii (a) $R\approx 4 \xi_0$ and (b) $R \approx 14 \xi_0$ plotted as functions of the disorder scattering rate $\Gamma$, obtained by fixing the distribution of impurities at density $\rho_\mathrm{imp} \approx 17 /\xi_0^2$ and by tuning the parameter $V_0$ in~\eqref{eq:imp_potential}. Larger colored dots denote the first excited states. Horizontal lines at selected values of $\Gamma$ denote deciles of the corresponding minigap value distributions obtained for 1000 random disorder realizations, with the longest lines denoting the medians. \textbf{Central column:} Density maps of the MZM wave functions' squared moduli in the antidots of radii (c)--(e) $R\approx 4 \xi_0$ and (f)--(h) $R\approx 14 \xi_0$, calculated (c), (f) without disorder, and (d), (e), (g), (h) with disorder with the scattering rate indicated in the plots. The color scale in (c) is common for all maps (c)--(h). \textbf{Right column:} Radial profiles of the MZM wave functions' squared moduli, drawn along the lines crossing the density maxima [shown as dotted lines in (c)--(h)] for the antidots of radii (i) $R \approx 4 \xi_0$ and (j) $R\approx 14 \xi_0$, obtained for several values of $V_0$ corresponding to evenly spaced values of $\Gamma$ corresponding to arrows in (a) and (b). Vertical spacing of (i) $2\times 10^6$ or (j) $1\times 10^6$ has been applied to the baselines of the curves to enhance clarity of the data. }
\label{fig:scan_dmu_alt}
\end{figure*}

First, we consider a fixed impurity density $\rho_\mathrm{imp} \xi_0^2 \approx 17$, which in our model corresponds to the fraction of 0.15 of the lattice sites being occupied by impurities, and two different antidot radii: $R= 1.26 \pi \xi_0 \approx 4 \xi_0$ and $R= 4.5 \pi \xi_0 \approx 14 \xi_0$. The smaller (larger) value of $R$ represents the regime in which the antidot area constitutes the minority (majority) of the area $\approx \pi(R+\pi \xi_0)^2$ occupied by the wave functions of states bound to the vortex core. The specific numerical values of $R$ are dictated by the convenience of numerical calculations. In Figs.~\ref{fig:scan_dmu_alt}(a) and~\ref{fig:scan_dmu_alt}(b), we present the calculated energy spectra for the two antidot sizes as dot plots, where the magnitude of the impurity potential $V_0$ is varied from $0$ to $15 \Delta_0$. For the chosen impurity density, this range corresponds to $\Gamma$ changing from $0$ to $4.3\Delta_0$. The introduction of the disorder potential $\delta \mu (\br)$ in the numerical model leads to a shift in the average of the chemical potential inside the antidot:
\begin{equation} \label{eq:muin_avg}
\langle \mu_\mathrm{in} \rangle = \mu_\mathrm{in,0} + \frac{1}{\pi R^2}\int_{r<R} \delta \mu (\br)\,d^2\br.
\end{equation}
We account for this effect by adjusting $\mu_\mathrm{in,0}$ such that $\langle \mu_\mathrm{in} \rangle=0$ in every calculation. 

To provide a statistical context to the spectra presented in Figs.~\ref{fig:scan_dmu_alt}(a) and~\ref{fig:scan_dmu_alt}(b), at selected values of $\Gamma$ deciles of distributions of the values of the minigap are shown as horizontal lines. This statistical data is derived from simulations of an ensemble of 1000 antidot systems, with varied impurity locations, and parameters $V_0$ and $\rho_\mathrm{imp}$ kept constant. The longest lines denote the median of each distribution. Histograms illustrating the distributions are shown in Figs.~\ref{fig:hist1}(a) and~\ref{fig:hist1}(b) in Appendix~\ref{app:stats}.

For both antidot radii, we find that for the weakest disorder the minigap decreases approximately linearly with increasing $\Gamma$, and attains a minimum at a certain critical value $\Gamma_c$, but does not reach zero due to the finite size of the antidot.  At the same time, the MZM wave function changes its distribution with growing $\Gamma$, as illustrated in Figs.~\ref{fig:scan_dmu_alt}(c)--\ref{fig:scan_dmu_alt}(j). In the case of a clean sample, the MZM wave function is smooth in the entire antidot area and distributed symmetrically along the polar axis around the antidot center. This is due to $\langle \mu_\mathrm{in} \rangle$ being fixed at the charge neutrality point of the TI surface Dirac spectrum. At $\Gamma>0$ the wave function, albeit still permeating the whole antidot, develops maxima at certain randomly located points. Figures~\ref{fig:scan_dmu_alt}(c)--\ref{fig:scan_dmu_alt}(h) show the wave function density maps for values of $\Gamma$ varied between $0$ and $0.5\,\Delta_0$, such that the minigap is open. The square moduli of the MZM wave function presented in the figure are equivalent to the local density of states (LDOS) profiles, as the MZM is the only state within the $0.1 \Delta_0$ range of zero energy. In Figs.~\ref{fig:scan_dmu_alt}(i) and~\ref{fig:scan_dmu_alt}(j), we present cross sections through the LDOS profiles for several more values of $\Gamma$. 

As illustrated most clearly in Fig.~\ref{fig:scan_dmu_alt}(b), after initially approaching zero, the minigap may increase with growing $\Gamma$ and subsequently oscillate with a decreasing amplitude, without affecting the presence of the MZM. However, for larger $\Gamma$ the MZM wave function may change its real-space distribution dramatically (see Fig.~\ref{fig:more_wavefs} in Appendix~\ref{app:stats}). We have verified that for the larger antidot radius the dependence of the minigap on $V_0$ has an oscillatory component for the majority of disorder realizations. In the specific case presented in Fig.~\ref{fig:scan_dmu_alt}(b), the oscillation is exceptionally prominent for such weak disorder. A more typical behavior of the minigap, observed in a simulation with a different disorder realization, is presented in Fig.~\ref{fig:scan_dmu_typical}.

\begin{figure}
\centering
\includegraphics[scale=1]{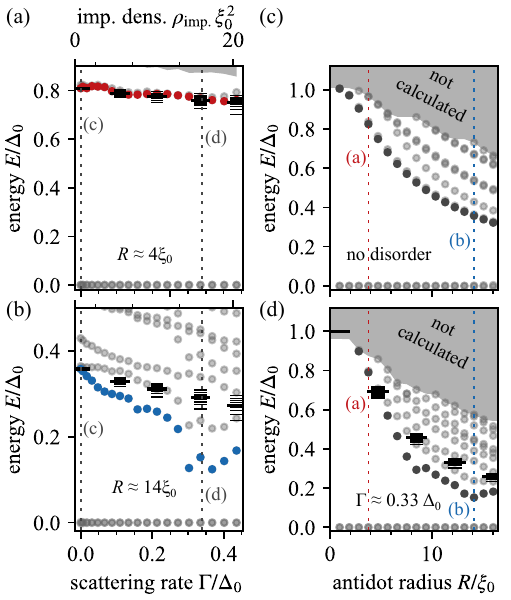}
\caption{\textbf{Left column:} Energy spectra of the vortex cores pinned inside the antidots of radii (a) $R \approx 4 \xi_0$ and (b) $R \approx 14 \xi_0$ as functions of the disorder scattering rate $\Gamma$, obtained by changing the impurity density $\rho_\mathrm{imp}$ with a fixed single-impurity potential magnitude $V_0=6 \Delta_0$. \textbf{Right column:} Analogous energy spectra as functions of the antidot radius (c) without disorder and (d) with disorder characterized by parameters $V_0=6 \Delta_0$ and $\rho_\mathrm{imp} \approx 17/\xi_0^2$. Larger colored or dark grey dots denote the first excited states. Horizontal lines at selected values of (a), (b) $\Gamma$ or (d) $R$ denote deciles of the corresponding minigap value distributions obtained for 1000 random disorder realizations, with the longest lines denoting the medians.  } 
 \label{fig:scan_rho_R}
\end{figure}

We complement the above results obtained for a fixed value of $\rho_\mathrm{imp}$ and variable $V_0$ with a series of calculations performed for fixed $V_0=6\Delta_0$ and changing $\rho_\mathrm{imp}$ with values ranging from 0 up to  $\rho_\mathrm{imp} \xi^2_0 \approx 20$, which corresponds to $\Gamma$ ranging from $0$ to $0.43 \Delta_0$.  The energy spectra calculated for the two antidot radii, $R\approx 4 \xi_0$ and $R\approx 14 \xi_0$, are presented in Figs.~\ref{fig:scan_rho_R}(a) and~\ref{fig:scan_rho_R}(b). The results for subsequent values of $\rho_\mathrm{imp}$ were obtained by successively adding impurity sites to the system, such that all impurity locations at a given value of $\rho_\mathrm{imp}$ are preserved in calculations for all larger $\rho_\mathrm{imp}$ values. We allowed the new impurity sites to fall both inside and outside of the antidot. At each step, $\mu_\mathrm{in,0}$ was adjusted according to Eq.~\eqref{eq:muin_avg} such that $\langle \mu_\mathrm{in} \rangle = 0$ at all times. Similarly to the case of fixed $\rho_\mathrm{imp}$ and variable $V_0$ described in Fig.~\ref{fig:scan_dmu_alt}, here we find that for small $\Gamma$ the decrease of the minigap with growing $\Gamma$ is approximately linear, albeit with some fluctuations which we attribute to the randomness of the process of increasing the impurity density. 

Our findings consistently indicate that the energy spectrum of the smaller antidot is less susceptible to disorder. The calculated dependence of the spectrum on the antidot radius is presented in Fig.~\ref{fig:scan_rho_R}(c) for the case of a clean sample, and Fig.~\ref{fig:scan_rho_R}(d) for the disordered one with a fixed disorder profile. In the case of no disorder, the minigap decreases monotonically, and at $R\gg  \xi_0$ becomes inversely proportional to $R$, while in the limit $R \rightarrow 0$ it saturates to a finite value. For the disordered case with $\Gamma/\Delta_0 \approx 0.33$, the decrease of minigap is not monotonic, which is due to the impurities being localized at random locations in the sample. As $R$ is increased, more and more impurity sites fall within the antidot area, and the mean value of $\delta \mu$ in the antidot fluctuates. At each step, $\mu_\mathrm{in,0}$ was adjusted such that $\langle \mu_\mathrm{in} \rangle = 0$. We find that the decrease of the minigap with growing $R$ is more significant in the disordered sample than in  the clean sample. Comparing Figs.~\ref{fig:scan_rho_R}(c) and~\ref{fig:scan_rho_R}(d), we conclude that antidots with the radius near $\pi \xi_0$ are not significantly affected by disorder and are thus favourable for potential applications in topological quantum computation.

In analogy to Figs.~\ref{fig:scan_dmu_alt}(a) and~\ref{fig:scan_dmu_alt}(b), the horizontal marks in Figs.~\ref{fig:scan_rho_R}(a), \ref{fig:scan_rho_R}(b), and~\ref{fig:scan_rho_R}(d) denote deciles of the statistical minigap distributions, calculated for several fixed values of (a), (b) $\rho_\mathrm{imp}$ or (d) $R$. The histograms illustrating these distributions are found in Appendix~\ref{app:stats} in Figs.~\ref{fig:hist1}(c) and~\ref{fig:hist2}(a).  

So far, we have adopted a fixed average chemical potential $\langle \mu_\mathrm{in} \rangle = 0$, although in a real device it would generally attain a different value determined by the specific properties of the materials comprising the heterostructure and its fabrication quality. However, it is our assumption that $\langle \mu_\mathrm{in} \rangle$ can be effectively controlled by an external electric field through tuning $\mu_\mathrm{in,0}$ in~\eqref{eq:muin_avg}. This could be achieved by introducing a gate terminal adjacent to the heterostructure in the vicinity of the antidot. Thus, in a given sample, one would ideally be able to optimize the minigap size by tuning the gate voltage. 
Additional tuning can be achieved by using multiple gates~\cite{PhysRevLett.130.106001}. 

\begin{figure}
\centering
\includegraphics[scale=1]{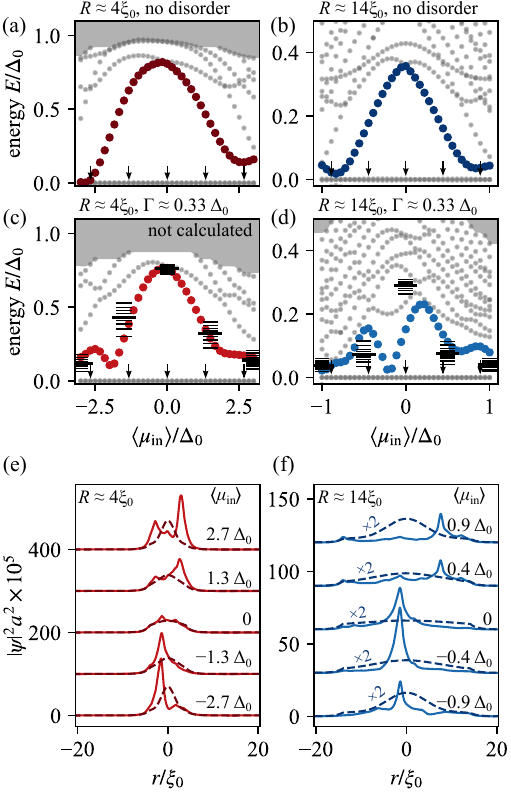}
\caption{
The influence of the chemical potential on the energy spectra of the vortex cores pinned inside the antidots of radii (a), (c), (e) $R\approx 4 \xi_0$ and (b), (d), (f) $R\approx 14 \xi_0$. The spectra are plotted as functions of the mean chemical potential in the antidot $\langle \mu_\mathrm{in} \rangle$, both (a), (b) in clean systems and (c), (d) in systems with disorder characterized by parameters $V_0=6 \Delta_0$ and $\rho_\mathrm{imp} \approx 17/\xi_0^2$, which at $\langle \mu_\mathrm{in} \rangle = 0$ corresponds to $\Gamma \approx 0.33 \Delta_0$. Larger colored dots denote the first excited states. Horizontal lines at selected values of $\langle \mu_\mathrm{in} \rangle$ denote deciles of the corresponding minigap value distributions obtained for 1000 random disorder realizations, with the longest lines denoting the medians. (e), (f) Radial profiles of the MZM wave functions
squared moduli, corresponding to states indicated by black arrows in (a)--(d). Dashed (solid) lines present the results for the clean (disordered) systems. To enhance the clarity of the data vertical spacing of (e) $1\times 10^7$ or (f) $3\times 10^6$ has been applied to the baselines of the curves. Dashed lines in (f) represent original data multiplied by a factor of 2. }
 \label{fig:scan_muin}
\end{figure}

In Figs.~\ref{fig:scan_muin}(a)--\ref{fig:scan_muin}(d) we present the energy spectra of the antidots of radii $R\approx 4 \xi_0$ and $R \approx 14 \xi_0$ plotted as functions of $\langle \mu_\mathrm{in} \rangle$, which is varied in the range of a few $\Delta_0$. Figures~\ref{fig:scan_muin}(a) and~\ref{fig:scan_muin}(b) present the results for clean samples, confirming that, in fact, $\langle \mu_\mathrm{in} \rangle/\Delta_0 \approx 0$ corresponds to the largest minigap. Note that the plotted spectra are not symmetric with respect to $\langle \mu_\mathrm{in} \rangle=0$, which is due to the chemical potential $\mu_\mathrm{out}$ outside the antidot having a non-zero value.

In the presence of disorder with $\Gamma \approx 0.33 \Delta_0$ (as calculated at $\langle \mu_\mathrm{in} \rangle=0$), the two samples with different radii respond differently to adjusting $\langle \mu_\mathrm{in} \rangle$. Figure~\ref{fig:scan_muin}(c) shows that for the antidot with $R \approx 4 \xi_0$ the optimal minigap again occurs at $\langle \mu_\mathrm{in} \rangle/\Delta_0 \approx 0$. On the contrary, the spectrum of the disordered antidot with $R \approx 14 \xi_0$ features states lying in the gap of the clean system. Due to their presence, $\langle \mu_\mathrm{in} \rangle/\Delta_0$ corresponding to the optimal minigap is shifted away from zero to a value dependent on the disorder realization.

For $|\langle \mu_\mathrm{in} \rangle|$ above a certain value, both in the clean and the disordered system, the minigap nearly closes, and the spectrum can feature densely spaced energy levels corresponding to trivial Caroli-de Gennes-Matricon (CdGM) states bound to the vortex core localized inside the antidot. This is consistent with scanning tunneling microscopy and spectroscopy (STM/STS) experiments with Abrikosov vortices in SC-TI heterostructures, where the bound states manifest themselves as an apparent splitting of the zero-bias peak at a certain distance from the vortex core~\cite{PhysRevLett.114.017001}.

Statistical distributions of the minigap value obtained for 1000 random disorder realizations at selected values of $\langle \mu_\mathrm{in} \rangle$ for both antidot sizes are shown in Figs.~\ref{fig:hist2}(b) and~\ref{fig:hist2}(c) in Appendix~\ref{app:stats}. 

Importantly, tuning the chemical potential $\langle \mu_\mathrm{in} \rangle$ also results in the change of the MZM wave function distribution. In a clean sample at $\langle \mu_\mathrm{in} \rangle = 0$, the MZM is almost evenly distributed throughout the antidot area and spills into the surrounding SC region with an exponentially decreasing amplitude. For nonzero $\langle \mu_\mathrm{in} \rangle$, however, the MZM density develops a distinct peak at the vortex core. The radial profiles of the MZMs in clean samples are shown as dashed lines in Figs.~\ref{fig:scan_muin}(e) and~\ref{fig:scan_muin}(f). In the disordered antidots case, on the other hand, the MZMs exhibit spatial fluctuations and peaks at random points, even at $\langle \mu_\mathrm{in} \rangle = 0$. Upon the variation of the chemical potential, the spatial profile of the wave function is altered, such that the existing peaks level off, and the new peaks emerge at different locations. The MZMs in disordered samples are represented by solid lines in Figs.~\ref{fig:scan_muin}(e) and~\ref{fig:scan_muin}(f). Such variation of the wave functions both in the clean and the disordered case can be attributed to the change of the Fermi momentum $k_F \approx \langle \mu_\mathrm{in} \rangle/(\hbar v_D) $ in the TI surface with the changing chemical potential, and the associated change of the density of states per unit area $\rho_\mathrm{TI} \approx | \langle \mu_\mathrm{in} \rangle| / (2 \pi \hbar^2 v_D^2) $. At larger $|\langle \mu_\mathrm{in} \rangle|$, states from a larger Fermi contour of the TI surface spectrum contribute to the formation of the MZM, allowing a tighter peak of the MZM amplitude near the vortex core in clean samples. In disordered samples the scattering from impurities obscures this effect. However, the change of the make-up of the Fermi contour with changing $\langle \mu_\mathrm{in} \rangle$ results in the MZM wave functions exhibiting different interference patterns. We propose that the evolution of the MZM wave function upon varying the gate voltage can be observed by means of the STM/STS method.

\begin{figure}
    \centering\includegraphics{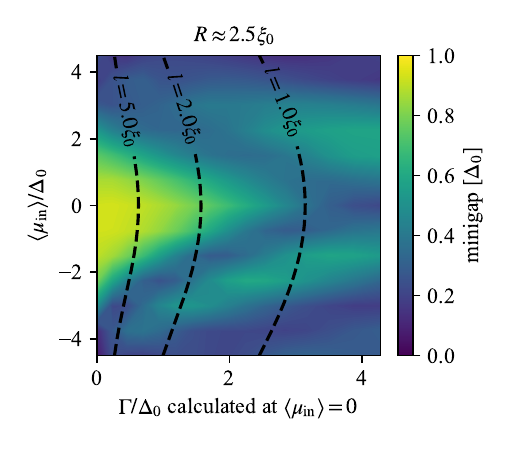}
    \caption{A color map of the minigap value for the antidot of radius $R \approx 2.5\xi_0$ as a function of the chemical potential inside the antidot $\langle \mu_\mathrm{in} \rangle$ and the disorder strength. The disorder strength is tuned by varying the parameter $V_0$ between $0$ and $15\Delta_0$ with a fixed distribution of impurities of density $\rho_\mathrm{imp} \approx 17/\xi_0^2$, and expressed in terms of the scattering rate $\Gamma$ calculated at $\langle \mu_\mathrm{in} \rangle = 0$. Black dashed lines are constant $\Gamma$ contours and  are labeled by the associated values of the mean free path $l=\hbar v_D/\Gamma$.}
    \label{fig:gap2d}
\end{figure}

The above results motivate a comprehensive study of the antidot with the smallest meaningful size, which is estimated by the radius of the core of an Abrikosov vortex $\approx \xi_0$. We expect that the disorder magnitudes allowed by our model have a minor effect on the spectra of such systems. Instead, we focus on the case of $R=0.8 \pi \xi_0 \approx 2.5 \xi_0$, which is closer to the theoretical limit than the previously investigated radii. We calculate the minigap as a function of both the chemical potential inside the antidot $\langle \mu_\mathrm{in} \rangle$ and the impurity potential strength $V_0$, for a fixed impurity density $\rho_\mathrm{imp} \approx 17/\xi_0^2$. The results are shown as a color map in Fig.~\ref{fig:gap2d}, with the disorder strength expressed as the scattering rate $\Gamma$ calculated for the chemical potential tuned to the Dirac point of the TI surface states ($\langle \mu_\mathrm{in} \rangle = 0$). 

Note that a uniform change of $\langle \mu_\mathrm{in} \rangle$, e.g., by applying gate voltage, corresponds to a vertical displacement in the plot in Fig.~\ref{fig:gap2d}. However, such a process would result in a change of measured $\Gamma$, since the density of states depends on $\langle \mu_\mathrm{in} \rangle$. For reference, contours denoting selected values of $\Gamma$ for arbitrary $\langle \mu_\mathrm{in} \rangle$ are included in Fig.~\ref{fig:gap2d} as black dashed lines and labeled with the associated value of the mean free path \eqref{eq:mfp}.

Noticeably, vertical line cuts of the data presented in Fig.~\ref{fig:gap2d} agree in qualitative terms with analogous spectra shown in Figs.~\ref{fig:scan_muin}(a)--\ref{fig:scan_muin}(d), obtained by fixing the disorder magnitude and varying $\langle \mu_\mathrm{in} \rangle$. Similarly, the horizontal line cut for $\langle \mu_\mathrm{in} \rangle=0$ is in agreement with the analogous data in Figs.~\ref{fig:scan_dmu_alt}(a) and~\ref{fig:scan_dmu_alt}(b). Very clearly, the minigap decreases both with growing $|\langle \mu_\mathrm{in} \rangle|$ and growing $\Gamma$, albeit with significant oscillations. Therefore, for a given disordered sample, a wide scan of $\langle \mu_\mathrm{in} \rangle$ has to be performed to find the configuration ensuring the maximal value of the minigap.

\section{Conclusions \label{sec:conclusions} } 
We performed a detailed numerical analysis of the TI-SC device with an antidot structure, using realistic parameters. We focused on how the minigap, which separates the zero-energy Majorana mode in the antidot from the first excited trivial CdGM state, depends on various factors, i.e., geometrical dimensions, Fermi energy, and disorder strength. We also examined how these factors affect the Majorana wave function profile inside the antidot. We considered a realistic correlated disorder model mimicking randomly distributed charged impurities. We established a relationship between impurity density, potential strength, and the corresponding scattering rate, see Fig.~\ref{fig:Gamma_vs_s}, which can be used to calibrate our disorder model
based on transport measurements. Our conclusions are therefore general and are not limited by the specific choice of disorder model.

Our results have implications for the current and future experiments on antidots in TI-SC devices, which are a promising platform for topological quantum computing. The minigap is the crucial parameter for such applications---it has to be larger than the temperature to enable fast and reliable quantum information processing and storage. We studied the dependence of the minigap on antidot radius in clean and disordered limits, as shown in Figs.~\ref{fig:scan_rho_R}(c) and~\ref{fig:scan_rho_R}(d), and found that small antidots with $R \leq 5\xi_0$ are fairly robust with respect to disorder. We also demonstrated that the electron density in the antidot strongly affects the minigap, with the minigap values peaking at or near the Dirac point $\langle \mu_\mathrm{in} \rangle=0$; see Figs.~\ref{fig:scan_muin}(a)--\ref{fig:scan_muin}(d) and~\ref{fig:gap2d}. These results emphasize the importance of being able to adjust the chemical potential inside the antidot using external gates to achieve the optimal minigap values in realistic experimental devices. In particular, we found that for some levels of disorder, the optimal value is obtained by tuning $\langle \mu_\mathrm{in} \rangle$  away from the charge neutrality point, see Fig.~\ref{fig:gap2d}.

We found that disorder has a negative effect on the minigap, which is the case in other proposed MZM platforms and is in qualitative alignment with earlier results of Ref.~\cite{PhysRevLett.130.106001}. The effects of disorder are shown in Figs.~\ref{fig:scan_dmu_alt}(a), \ref{fig:scan_dmu_alt}(b), and~\ref{fig:gap2d}: the minigap decreases in an oscillatory manner, nearly closing at certain critical values $\Gamma_c$. However, we also found that by tuning electron density inside the antidot, one is able to tune away from the local minima of the minigap, as demonstrated in Figs.~\ref{fig:scan_muin}(d) and~\ref{fig:gap2d}.

Our findings are supplemented by a basic statistical analysis of the distribution of minigap values in random disorder realizations. For the strongest investigated disorder in the antidot of radius $R\approx 14\xi_0$, which falls within the regime $\xi_0 \ll R$, we approximately recover the theoretical distribution of the minigap predicted by random matrix theory~\cite{PhysRevLett.130.106001} (see Appendix~\ref{app:stats}).

For the smallest investigated antidot size with $R\approx 2.5\xi_0$, we found that the scattering rate as high as $\Gamma \approx 4.5\Delta_0$ allows a minigap of about $0.5 \Delta_0$. This value is comparable to the largest minigap values observed for the antidot of radius $R\approx 4\xi_0$ at similar $\Gamma$; mesoscopic fluctuations of the minigap allow for large minigaps in certain disorder realizations, as indicated by the broad distribution of minigaps (see Fig.~\ref{fig:hist1} in Appendix~\ref{app:stats}).

For typical TIs, such as $(\text{Bi}_{0.4}\text{Sb}_{0.6})_2\text{Te}_3$ (BST)~\cite{2021arXiv211001039R}  and $\text{BiSbTeSe}_2$ (BSTS)~\cite{2014NatPh..10..956X}, the reported values of the mean free path, extracted from Hall measurements, are $l\approx15\,\mathrm{nm}$ and $l\approx17\,\mathrm{nm}$, respectively. 
Taking into account the reported Dirac velocities, $v_D \approx 4 \times 10^5\, \text{m}/ \text{s}$ for BST and $v_D \approx 3 \times 10^5\, \mathrm{m}/ \mathrm{s}$ for BSTS, we estimate the corresponding scattering rates, $\Gamma = \hbar v_D / l$, to be $\Gamma\approx 17.5\,\mathrm{meV}$ for BST and $\Gamma\approx 11.6\,\mathrm{meV}$ for BSTS, respectively. Hence, we have found that for antidots of radii $\sim \pi \xi_0$ or larger, the scattering rate in state of the art TIs may be too high to exhibit a detectable minigap, even for large-gap superconductors, such as Nb. Thus, cleaner devices or smaller-sized antidots would be required to achieve a sizable minigap.
For example, taking BSTS compound as a TI, Nb as a SC ($\Delta_0\approx 1\,\mathrm{meV}$), and the antidot of the radius $R \approx 2.5 \xi_0$, the electronic mean free path has to be increased to $l>45 \,\mathrm{nm}$ from the current $l\approx17 \,\mathrm{nm}$ to achieve a meaningful value of the minigap. Nevertheless, our findings are encouraging in the sense that for reasonably small-sized antidots ($R<5\xi_0$), the disorder strength required for the closing of the minigap corresponds to a scattering rate that is several times  higher than the gap of a clean system. In addition, the minigap can be reopened by gating, as illustrated in Fig.~\ref{fig:gap2d}. While our use of a low-energy effective  model prevents us from doing simulations at stronger disorder strength and/or higher chemical potential, 
we envision that the diagram in Fig.~\ref{fig:gap2d} continued for higher values of $\Gamma$ involves further oscillations of the minigap, and includes more areas where a significant minigap magnitude can be found. Moreover, the distribution of the minigap values can be quite broad, as shown in Fig.~\ref{fig:hist1}(b), suggesting that in a series of samples (or in a single sample with multiple gates~\cite{PhysRevLett.130.106001}), it is likely possible to find a disorder configuration with the minigap significantly exceeding the ensemble average.

\section*{Acknowledgments } 
The work of R.R. is supported by the Foundation for Polish Science through the International Research Agendas program co-financed by the European Union within the Smart Growth Operational Programme. 
This material is based upon work supported by the U.S. Department of Energy, Office of Science, National Quantum Information Science Research Centers, Quantum Science Center. 
Part of the work by A.K. was supported by the U.S. Department of Energy, Office of Science, Basic Energy Sciences, Materials Science and Engineering Division, including the grant of computer time at the National Energy Research Scientific Computing Center (NERSC) in Berkeley, California. This part of research was performed at the Ames National Laboratory, which is operated for the U.S. DOE by Iowa State University under Contract No. DE-AC02-07CH11358.

\appendix

\section{Validity of the lattice model} \label{app:validity}
\subsection{Fermi contour warping}
For small $a\bk$ the Hamiltonian~\eqref{eq:TI_surf_Hamiltonian} reduces to a simple Dirac Hamiltonian
\begin{equation} \label{eq:Dirac_Ham}
    h_\mathrm{TI} \approx \lambda a ( s_x k_x + s_y k_y) - \mu_0
\end{equation}
characterized by a circular Fermi contour. In the lattice model, at the energy at which the terms of higher order in $a\bk$ become significant, the Fermi contour warps. We estimate this energy threshold by comparing the first- and third-order terms of the expansion of $\sin(a |k|)$,
\begin{equation}
     a |k| \gg \frac{a^3 |k|^3}{6} \quad \Rightarrow \quad \left|\frac{E}{\Delta_0} \right| \ll \sqrt{ 6 } \frac{\pi \xi_0}{a} ,
\end{equation}
where we used $|E|=\lambda a |k|$. In our model, $a = 0.03 \pi \xi_0$, and thus the Fermi contour warping can be neglected if $|E|\ll 81.7 \Delta_0$.

\subsection{TRS breaking}
Near $\Gamma$, the effect of the TRS breaking terms in the Hamiltonian~\eqref{eq:TI_surf_Hamiltonian} is given by $\frac{m a^4}{8}(k_x^4+k_y^4) s_z $. We compare the physical Dirac term in~\eqref{eq:Dirac_Ham} to the magnetic perturbation:
\begin{equation}
    |\lambda a k| \gg  \left|\frac{m a^4}{8} k^4 \right| \quad \Rightarrow \quad \left|\frac{E}{\Delta_0} \right| \ll  2  \frac{ \pi \xi_0 }{a  } \sqrt[3]{ \left|\frac{\lambda}{ m } \right|}.
\end{equation}
For the parameter values used in our lattice model, the effect of TRS breaking terms near $\Gamma$ is negligible if $|E| \ll 58.3 \Delta_0$.

Furthermore, we note that the magnetic gap at $\mathrm{X}$ and $\mathrm{Y}$ points in the Brillouin zone is $4|m|=200 \Delta_0$, and at $\mathrm{M}$ it is $8|m| = 400 \Delta_0$. Our analysis is confined within these energy gaps.

\section{Estimation of the scattering rate for the disorder model \eqref{eq:delta_mu}-\eqref{eq:imp_potential}}\label{app:scattering}
The scattering rate $1/\tau_\bk$ can be estimated using Fermi's golden rule,
\begin{equation} \label{eq:Fermi_golden}
        \frac{1}{\tau_\bk} = \frac{2 \pi}{\hbar} \sum_{\bk'} |\braket{\bk|\delta \mu|\bk'}|^2 \delta(\epsilon_\bk - \epsilon_{\bk'}),
\end{equation}
where $\ket{\bk}$ and $\ket{\bk'}$ are the eigenstates of the 2D Dirac Hamiltonian~\eqref{eq:Dirac_Ham} at zero energy. Without loss of generality, we choose $\mu_0>0$. The square modulus of the matrix element of $\delta\mu$, as defined in~\eqref{eq:delta_mu}, is
\begin{multline} \label{eq:scatt_ampl}
    |\braket{\bk|\delta \mu|\bk'}|^2 = \frac{V_0^2}{\mathcal{N}} \frac{4 \pi^2 \sigma^4}{\Omega^2}   \\ \times
   \exp \left[  - \sigma^2 (\bk' - \bk )^2
    \right] \frac{1+\cos{ (\phi-\phi')}}{2}  \\  \times
   \sum_{n=1}^{2 N_C} \sum_{m=1}^{2 N_C} \left( \eta_n \eta_m
        \exp \left[   i (\bk' - \bk ) \cdot (\br^C_n - \br^C_m)
    \right]
    \right),
\end{multline}
where $\Omega$ is the surface area, $\phi^{(\prime)} $ is the polar angle in the reciprocal space, and 
\begin{equation}
\eta_n = 
    \begin{cases}
        1 & \textrm{for } n = 1,2,
        \ldots,N_C,\\
        -1 & \textrm{for } n = N_C+1,N_C+2,
        \ldots,2N_C.\\
    \end{cases}
\end{equation}
In the double sum in~\eqref{eq:scatt_ampl} the terms with $m=n$ sum to $2 N_C$. Assuming the uniform probability distribution of $\br^C_i$, the disorder average of the exponents in the remaining terms, in the limit of infinite $\Omega$, is
\begin{equation}
    \overline{ \exp \left[   i (\bk' - \bk ) \cdot (\br^C_n - \br^C_m) \right] } \approx \delta_{\bk\bk'}.
\end{equation}
Thus, for $\bk\neq\bk'$
\begin{multline}
        \overline{ |\braket{\bk|\delta \mu|\bk'}|^2 } \approx \frac{V_0^2}{\mathcal{N}} \frac{\rho_\mathrm{imp} 4 \pi^2 \sigma^4}{\Omega} \\ \times \frac{1+\cos{ (\phi-\phi')}}{2} \exp \left[  - \sigma^2 (\bk' - \bk )^2
    \right],
\end{multline}
where $\rho_\mathrm{imp} = 2 N_C/\Omega$ is the impurity density per unit area. In the limit of the infinite surface area, the summation in~\eqref{eq:Fermi_golden} is replaced with the integration, and the $\bk=\bk'$ term becomes negligible. The integral evaluated for small $\sigma |\bk|$ thus yields
\begin{equation}  \label{eq:scatter2}
    \frac{1}{\tau_\bk} \approx  
     \frac{2 \pi^2}{\hbar}  \frac{|\mu_0|}{\hbar^2 v_D^2} \frac{  V_0^2 }{ \mathcal{N}} \sigma^4 \rho_\mathrm{imp}  
    \left[1  - \left( \frac{|\mu_0|\sigma}{\hbar v_D} \right)^2  \right].
\end{equation}
Here we used the fact that the Fermi momentum satisfies $|\bk| = |\mu_0/(\hbar v_D)|$ and that the density of states per unit area is $|\mu_0| / (2\pi \hbar^2 v_D^2)$.

As our primary interest lies in the case of $\mu_0=0$, we replace the chemical potential in~\eqref{eq:scatter2} with appropriate averages of $|\delta\mu|$. Per the central limit theorem, at any position $\br$ the probability density function of $\delta\mu(\br)$ is given by the normal distribution
\begin{equation} \label{eq:dm_distibution}
    f(\delta \mu) = \frac{1}{\sqrt{2\pi} s} \exp \left(- \frac{\delta \mu^2}{ 2 s^2  } \right),
\end{equation}
where
\begin{equation}
    s^2 = \mathrm{Var}[\delta\mu(\br)] \approx 2 N_C \mathrm{Var}[V(\br)] \approx \rho_\mathrm{imp} \frac{V^2_0 \pi\sigma^2 }{\mathcal{N}}.
\end{equation}
It is straightforward to show that the averages in the normal distribution are
\begin{equation}
\langle |\delta \mu| \rangle_{\delta \mu} = 
 \sqrt{ \frac{2}{\pi}} s, \quad
\langle |\delta \mu|^3 \rangle_{\delta \mu} = 
2\sqrt{ \frac{2}{\pi}} s^3,
\end{equation}
from which it follows that
\begin{equation} \label{eq:scatter_final}
    \left\langle \frac{1}{\tau_\bk} \right\rangle_{\delta \mu} \approx  
     \frac{2 \sqrt{ 2} \pi^2}{\hbar}  \frac{  |V_0|^3 }{ \hbar^2 v_D^2 \mathcal{N}^{\frac{3}{2}}} \sigma^5 \rho_\mathrm{imp}^{\frac{3}{2}}  
    \left(1  - 2 \pi \frac{\rho_\mathrm{imp} V^2_0 \sigma^4 }{\mathcal{N}\hbar^2 v_D^2}  \right).
\end{equation}

\begin{figure}
    \centering
    \includegraphics{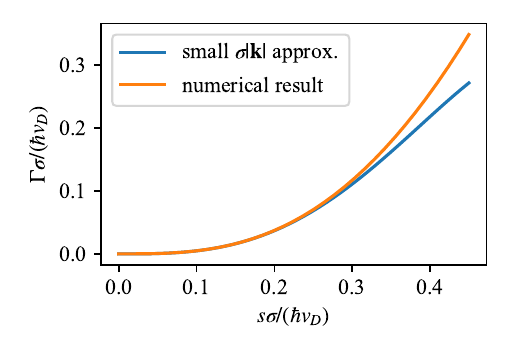}
    \caption{The dependence of the scattering rate $\Gamma$ on the variance of the disorder potential $s$, both in units of $v_D/\sigma$, evaluated for average chemical potential $\mu_0 = 0$. Approximation for small impurity potential fluctuations given by Eq.~\eqref{eq:scatter_rate} (blue curve) and the true dependence evaluated by means of numerical integration (orange curve).}
    \label{fig:Gamma_vs_s}
\end{figure}

In~\eqref{eq:scatter_rate}, we express the estimated scattering rate $\Gamma = \hbar \langle 1/\tau_\bk \rangle_{\delta\mu}$ in terms of $s$. Multiplying both sides of~\eqref{eq:scatter_rate} by $\sigma/(\hbar v_D)$, we find that $\Gamma \sigma/(\hbar v_D)$ depends only on one dimensionless parameter $s \sigma/(\hbar v_D)$. This dependence is plotted in Fig.~\ref{fig:Gamma_vs_s}.

For larger fluctuations $\delta \mu$ and for $\mu_0\neq 0$, the approximation expressed in~\eqref{eq:scatter_rate} and~\eqref{eq:scatter2} is insufficient. To describe these cases, we evaluate the integral exactly,
\begin{multline} \label{eq:exact_tauk}
    \frac{1}{\tau_\bk} =\frac{2 \pi^2}{\hbar}   \frac{|\mu_0| }{\hbar^2 v_D^2 } \frac{V_0^2}{\mathcal{N}} \rho_\mathrm{imp} \sigma^4 
    \exp \left(  - 2\frac{\mu_0^2 \sigma^2}{\hbar^2 v_D^2}   \right) \\
    \left[I_{0}\left(2\frac{\mu_0^2 \sigma^2}{\hbar^2 v_D^2} \right) + I_{1}\left(2\frac{\mu_0^2 \sigma^2}{\hbar^2 v_D^2} \right)\right],
\end{multline}
where $I_\alpha(x)$ are modified Bessel functions of the first kind. Then, we substitute $\mu_0 \rightarrow \mu_0 + \delta \mu$ and calculate numerically averages of $1/\tau_\bk$ with respect to the probability distribution~\eqref{eq:dm_distibution}. The scattering rate obtained in this way is compared to the approximate formula~\eqref{eq:scatter_rate} in Fig.~\ref{fig:Gamma_vs_s}, for the special case of $\mu_0 = 0$. 

\section{Minigap statistics of an ensemble of disordered antidot systems} \label{app:stats}
To complement the analyses presented in the main text, we performed additional calculations of the energy spectra of antidot systems for selected sets of parameters that correspond to those used in Sec.~\ref{sec:results}. For each parameter set, we examined an ensemble of 1000 different realizations of random disorder, allowing for a statistically robust evaluation of disorder effects on the minigap values in these systems. As we will show below, our results are qualitatively consistent with the findings of Ref.~\cite{PhysRevLett.130.106001}, which considered the regime of $\xi_0 \ll R$. 

\begin{figure*}
    \centering
\includegraphics[scale=1]{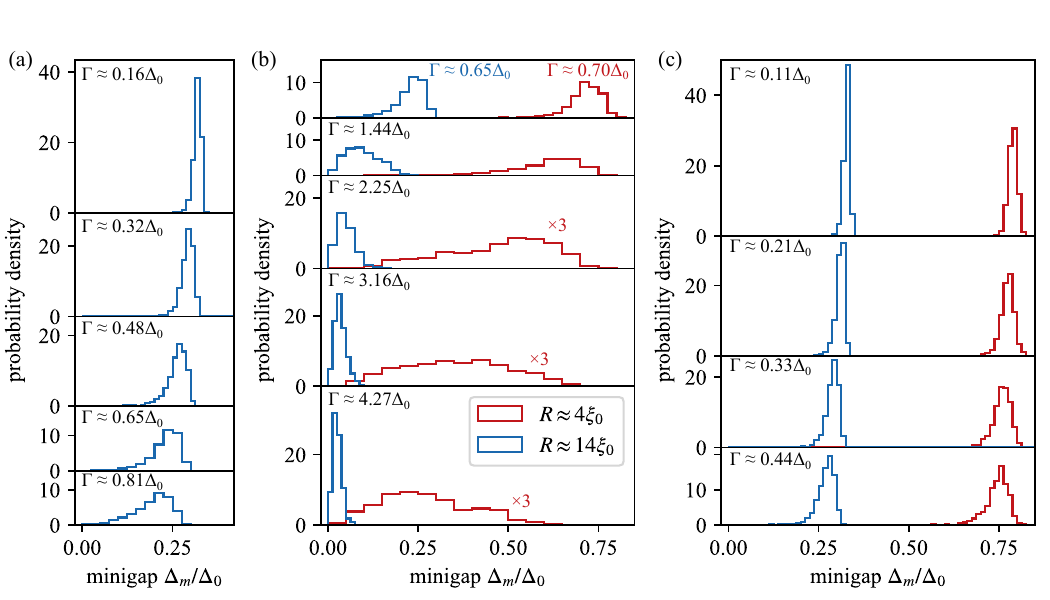}
\caption{ Histograms representing the statistical distributions of minigap $\Delta_m$ values in antidot systems, calculated numerically for an ensemble of 1000 random disorder realizations, for selected system parameters: (a), (b) constant impurity density $\rho_\mathrm{imp} \approx 17 /\xi_0^2$, and varying impurity potential strengths $V_0$, (c) constant $V_0 = 6\Delta_0$, with varying impurity densities $\rho_\mathrm{imp}$. Red (blue) curves correspond to antidot radius $R\approx 4 \xi_0$ ($R \approx 14 \xi_0$). The subpanel labels indicate the scattering rate values for the depicted curves. All results were obtained for chemical potential $\langle \mu_\mathrm{in} \rangle = 0$. The histograms are normalized to show probability density and the bin widths are set at 0.0125, 0.025, or 0.05, based on the spread of the distribution. Data shown in (a) and (b) correspond to  Figs.~\ref{fig:scan_dmu_alt}(a) and~\ref{fig:scan_dmu_alt}(b) in the main text. Data shown in (c) correspond to Figs.~\ref{fig:scan_rho_R}(a) and~\ref{fig:scan_rho_R}(b).
} \label{fig:hist1}
\end{figure*}

Figure~\ref{fig:hist1} presents histograms of the minigap $\Delta_m$ values in antidot systems. These distributions are also presented in the form of decile marks in Figs.~\ref{fig:scan_dmu_alt}(a), \ref{fig:scan_dmu_alt}(b), \ref{fig:scan_rho_R}(a), and \ref{fig:scan_rho_R}(b). Figure~\ref{fig:hist1}(a) shows results for the largest antidot radius, $R\approx 14 \xi_0$, at a fixed impurity density $\rho_\mathrm{imp} \approx 17 /\xi_0^2$, and different impurity potential strengths $V_0$, corresponding to scattering rates  $\Gamma < \Delta_0$. In this regime, the distributions exhibit negative skewness, reflecting the oscillatory dependence of the minigap on $\Gamma$. Mean $\Delta_m$ decreases at an approximate rate of $0.2$ units per unit increase in $\Gamma$. Subsequent histograms for $\Delta_0 < \Gamma < 4.5\Delta_0$, are shown in blue in Fig.~\ref{fig:hist1}(b). In this range of $\Gamma$, mean $\Delta_m$ continues to decrease towards zero at a reduced rate, with skewness of the distribution changing to positive between $\Gamma$ values $0.81 \Delta_0$ and $1.44 \Delta_0$. This positive skewness is attributed to the repulsion of the first excited state from the MZM. The standard deviation of the minigap distribution initially increases up to $\Gamma \approx 0.8$, and then decreases.

Analogous data for the smaller antidot with $R\approx 4 \xi_0$ is shown in red in Fig.~\ref{fig:hist1}(b) for $\Gamma < 4.5 \Delta_0$. In this case, mean $\Delta_m$ decreases approximately linearly across the examined range at a rate of approximately $0.15$ units per unit of $\Gamma$. The distributions are negatively skewed for $\Gamma \lesssim 3 \Delta_0$ and become positively skewed above this value. The standard deviation increases initially, peaks at $\Gamma \approx 2.2 \Delta_0$, and subsequently decreases. We observe that the agreement of the histograms with Ref.~\cite{PhysRevLett.130.106001} is better for the larger antidot and stronger disorder, which corresponds to $\xi_0, l\ll R$.

\begin{figure}
    \centering
    \includegraphics{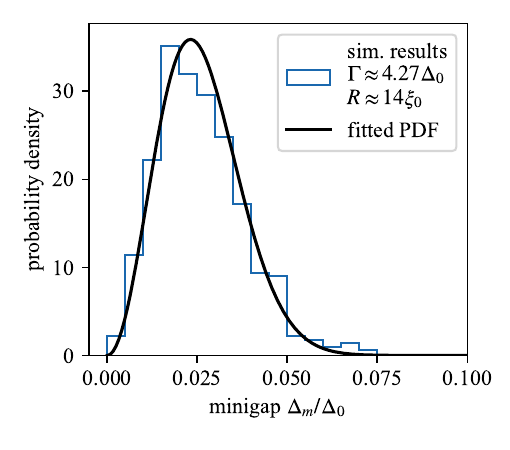}
    \caption{Probability density function~\eqref{eq:pdf} (black curve) fitted to the histogram of the minigap values calculated numerically for an ensemble of 1000 random disorder realizations. The presented histogram depicts the data already shown in the bottom row of Fig.~\ref{fig:hist1}(b), using a smaller bin width (0.005).  }
    \label{fig:fitted}
\end{figure}

For the strongest investigated disorder, the minigap distribution in the larger antidot follows approximately the analytical distribution
\begin{equation} \label{eq:pdf}
    P(\Delta_m) = \sqrt{  \frac{2}{\pi} } \frac{\Delta^2_m}{S^3} \exp \left(-\frac{\Delta^2_m}{2 S^2}\right),
\end{equation}
derived from random matrix theory~\cite{PhysRevLett.130.106001}, where the parameter $S$ is proportional to the average minigap $\overline{\Delta_m}=\sqrt{8/\pi}S$. Figure~\ref{fig:fitted} shows the graph of the analytical probability distribution function~\eqref{eq:pdf} fitted to the histogram of the simulated minigap values for $R\approx 14 \xi_0$, $\Gamma \approx 4.26 \Delta_0$, and $\langle \mu_\mathrm{in} \rangle = 0$. The fitting was performed by matching the calculated disorder minigap average $\overline{\Delta_m} = 0.026 \Delta_0$ to the theoretical one.

Figure~\ref{fig:hist1}(c) depicts histograms calculated in the same manner as above, but with a constant impurity potential strength $V_0 = 6 \Delta_0$, and varying impurity density $\rho_\mathrm{imp}$, corresponding to $\Gamma <0.5 \Delta_0$. We find that for both investigated antidot radii, the evolution of the minigap distributions is consistent with the findings of the analysis with constant $\rho_\mathrm{imp}$ and varying $V_0$. In particular, the rates of the decrease of mean $\Delta_m$ with growing $\Gamma$ are in approximate agreement with the values indicated in the above paragraphs. 

\begin{figure}
    \centering
\includegraphics[scale=1]{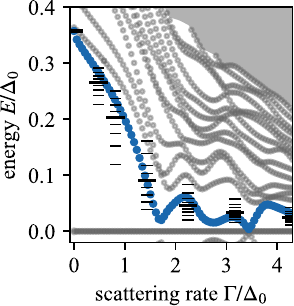}
\caption{
Energy spectra of the vortex core pinned inside the antidot of radius $R \approx 14 \xi_0$ plotted as a function of the disorder scattering rate $\Gamma$, obtained by fixing the distribution of impurities at density $\rho_\mathrm{imp} \approx 17 /\xi_0^2$ and by tuning the parameter $V_0$ in~\eqref{eq:imp_potential}. Larger colored dots denote the first excited states. Horizontal lines at selected values of $\Gamma$ denote deciles of the corresponding minigap value distributions obtained for 1000 random disorder realizations, with the longest lines denoting the medians. The figure presents data obtained with a disorder realization different than in Fig.~\ref{fig:scan_dmu_alt}(b).   } \label{fig:scan_dmu_typical}
\end{figure}

\begin{figure*}
\centering
\includegraphics[scale=1]{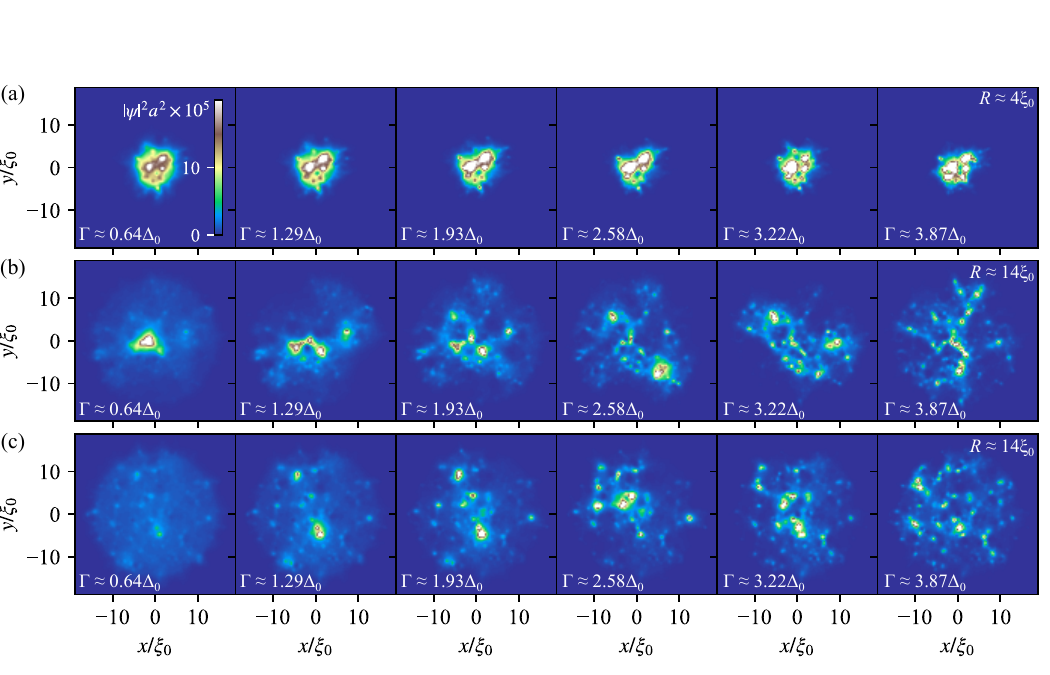}
\caption{
Density maps of the MZM wave functions' squared moduli in the antidots of radii (a) $R\approx 4 \xi_0$ and (b), (c) $R\approx 14 \xi_0$, the latter calculated for two different disorder realizations. The color scale is common for all maps. The data shown in (a) and (b) correspond to Fig.~\ref{fig:scan_dmu_alt}(a) and Fig.~\ref{fig:scan_dmu_alt}(b), respectively. The data in (c) correspond to Fig.~\ref{fig:scan_dmu_typical}.  } \label{fig:more_wavefs}
\end{figure*}

To supplement Fig.~\ref{fig:scan_dmu_alt}(b), in Fig.~\ref{fig:scan_dmu_typical} we present a plot of the energy spectrum of the antidot system versus the impurity potential strength, made for the antidot with radius $R\approx 14 \xi_0$, with a different disorder realization than the one used in the main text. This realization is selected to represent a more typical behavior of the minigap with growing disorder strength. Moreover, in Fig.~\ref{fig:more_wavefs} we show a selection of MZM wave function density maps for several selected values of the impurity potential strength $V_0$, corresponding to scattering rates between $\Gamma\approx 0.64 \Delta_0$ and $\Gamma \approx 3.87 \Delta_0$. Figure~\ref{fig:more_wavefs}(a) displays MZMs in the smaller antidot with $R\approx 4 \xi_0$, while Figs.~\ref{fig:more_wavefs}(b) and~\ref{fig:more_wavefs}(c) show MZMs for the antidot with $R\approx 14 \xi_0$. The data shown in Fig.~\ref{fig:more_wavefs}(b) was obtained for the disorder realizations for which the minigap oscillates quickly as a function of $\Gamma$, as presented in Fig.~\ref{fig:scan_dmu_alt}(b). On the other hand, the data shown in Fig.~\ref{fig:more_wavefs}(c) was obtained for the disorder realizations for which the minigap exhibits a more typical behavior depicted in Fig.~\ref{fig:scan_dmu_typical}.

\begin{figure*}
    \centering
\includegraphics[scale=1]{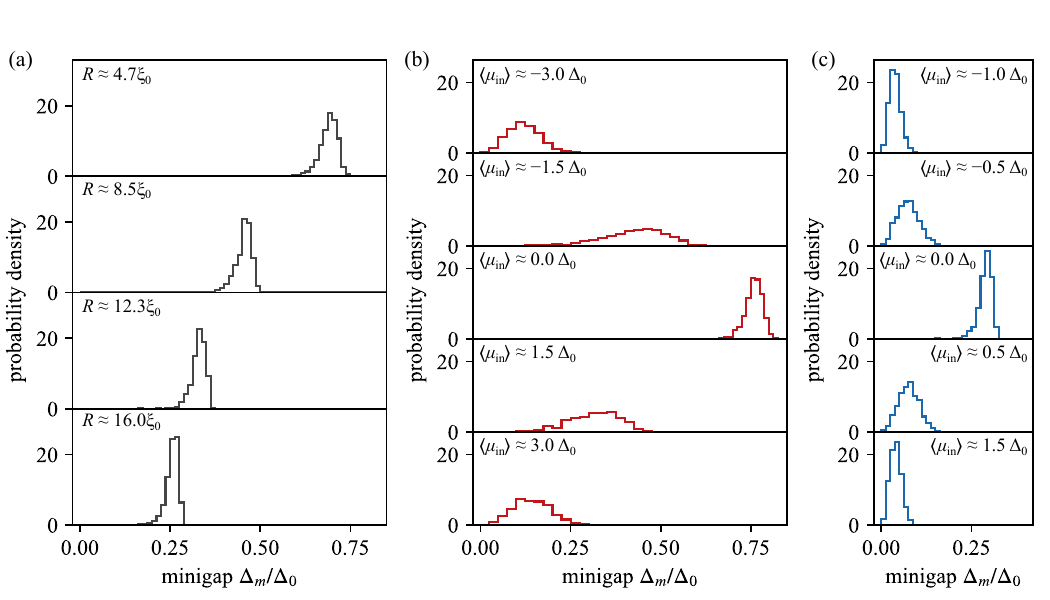}
\caption{
Histograms representing the statistical distributions of minigap $\Delta_m$ values in antidot systems, calculated numerically for an ensemble of 1000 random disorder realizations, for selected system parameters: constant impurity density $\rho_\mathrm{imp} \approx 17 /\xi_0^2$ and potential strength $V_0 = 6\Delta_0$ (corresponding to $\Gamma \approx 0.33 \Delta_0$), with (a) varying antidot radius $R$ (indicated in the subpanel labels) and constant chemical potential $\langle \mu_\mathrm{in} \rangle = 0$ , and (b), (c) varying chemical potential $\langle \mu_\mathrm{in} \rangle$ (indicated in the subpanel labels) and constant antidot radius (b) $R\approx 4 \xi_0$ or (c) $R\approx 14 \xi_0$. The histograms are normalized to show probability density and the bin widths are set at 0.0125, or 0.025, based on the spread of the distribution. Data shown in (a) correspond to Fig.~\ref{fig:scan_rho_R}(d) in the main text. Data shown in (b), (c) correspond to Fig.~\ref{fig:scan_muin}.} 
\label{fig:hist2}
\end{figure*}

Finally, Fig.~\ref{fig:hist2} presents histograms of the minigap values in antidot systems, corresponding to the decile marks shown in Figs.~\ref{fig:scan_rho_R}(d), \ref{fig:scan_muin}(c), and~\ref{fig:scan_muin}(d). The data in Fig.~\ref{fig:hist2} was obtained with a fixed impurity density $\rho_\mathrm{imp} \approx 17 /\xi_0^2$ and impurity potential strength $V_0 = 6 \Delta_0$, corresponding to scattering rate $\Gamma \approx 0.33 \Delta_0$. In Fig.~\ref{fig:hist2}(a), we show histograms generated for four different antidot radii. As expected, the mean minigap value drops with increasing radius. Figures~\ref{fig:hist2}(b) and~\ref{fig:hist2}(c) depict histograms of the minigap calculated at different values of chemical potential $\langle \mu_\mathrm{in} \rangle$ inside the antidots with radii $R\approx 4\xi_0$ and $R\approx 14 \xi_0$. In both cases, mean $\Delta_m$ decreases as $\langle \mu_\mathrm{in} \rangle$ is tuned away from charge neutrality point. As the minigap approaches zero, repulsion of the first excited state from the MZM can be observed, manifesting itself in the positive skewness of the minigap distribution, in analogy to the data presented in Figs.~\ref{fig:hist1}(a) and~\ref{fig:hist1}(b).

\bibliography{refs}

\end{document}